# ON THE keV STERILE NEUTRINO SEARCH IN ELECTRON CAPTURE

P.E. Filianin[1,2,*], K. Blaum[3], S.A. Eliseev[3], L. Gastaldo[4],
Yu.N. Novikov[1,2], V.M. Shabaev[1], I.I. Tupitsyn[1], J. Vergados[5]

[1]*St.Petersburg State University, St.Petersburg, Russia;*
[2]*Petersburg Nuclear Physics Institute, Gatchina, Russia;*
[3]*Max-Planck-Institut für Kernphysik, Heidelberg, Germany;*
[4]*Kirchhoff-Institute for Physics, Ruprecht-Karls-University, Heidelberg, Germany;*
[5]*University of Ioannina, Ioannina, Greece*
*corresponding author: E-mail: pavelfilyanin@yandex.ru

**Abstract.** A joint effort of cryogenic microcalorimetry (CM) and high-precision Penning-trap mass spectrometry (PT-MS) in investigating atomic orbital electron capture (EC) can shed light on the possible existence of heavy sterile neutrinos with masses from 0.5 to 100 keV. Sterile neutrinos are expected to perturb the shape of the atomic de-excitation spectrum measured by CM after a capture of the atomic orbital electrons by a nucleus. This effect should be observable in the ratios of the capture probabilities from different orbits. The sensitivity of the ratio values to the contribution of sterile neutrinos strongly depends on how accurately the mass difference between the parent and the daughter nuclides of EC-transitions can be measured by, e.g., PT-MS. A comparison of such probability ratios in different isotopes of a certain chemical element allows one to exclude many systematic uncertainties and thus could make feasible a determination of the contribution of sterile neutrinos on a level below 1%. Several electron capture transitions suitable for such measurements are discussed.

## 1. Introduction.

Sterile neutrinos are not part of the simple Standard Model of elementary particles and hence their discovery would point to new physics [1]. Recent explorations of the short baseline (reactor and β-decay) and the long baseline (accelerator) neutrino oscillation anomalies [2] hint at the possible existence of light sterile neutrinos in the 1 eV mass region [3]. Cosmological and galactic observations have pointed out that heavy sterile neutrinos in the keV mass region could be good warm dark matter (WDM) candidates [4, 5].

The neutrino flavor states can be considered a mixture of three active neutrino and one sterile neutrino eigenstates with the corresponding unitary mixing matrix. Since all active neutrinos are much lighter than the sterile one we can consider in the keV scale a simplified version of the mixture of one active mass eigenstate with $m_i < 1$ eV and one sterile neutrino mass eigenstate with $m_s$ in the keV scale:

$$|\nu_e\rangle = \cos\theta |\nu_i\rangle + \sin\theta |\nu_s\rangle, \qquad (1)$$

where θ is the mixing angle between the active $\nu_i$ and the sterile $\nu_s$ state. Note that this is true even if $\nu_i$ represents the sum of all standard neutrinos if they are unresolved. Thus $\sin^2\theta \equiv U_{e4}^2$ gives the contribution of the sterile neutrino to the probability of the corresponding processes.

It is worth noting here that this formalism resembles the approach used for the explanation of the presumable existence of so called Simpson's 17 keV neutrinos [6, 7]. The existence of 17 keV neutrinos has been suggested as a result of the observation (though eventually mistaken) of a kink in continuous beta-spectra. A proposal to search for such massive neutrinos in EC has also been formulated in [8].

From the seesaw model one can estimate the mixing angle as $m_D / M$, where $m_D$ is the Dirac mass and $M$ is the sterile right-hand Majorana neutrino mass. Assuming $m_D \approx 0.1$ eV and $M \approx 1$ keV one can find for the mixing angle a value of $10^{-4}$, which is considered large enough to produce the necessary amount of DM [9]. Cosmology [10] yields for sinθ a value of $10^{-4}$, which agrees with the above considerations. It is worth noting, however, that all mixing angle values in vogue are model dependent. Until now, there has been no direct indication of sterile neutrinos. The existence of cosmic sterile neu-trinos can be manifested by an observation of a monochromatic X-ray line released from the radiative decay of sterile neutrinos $\nu_s$ into active ones $\nu_a$: $\nu_s \to \nu_a + \gamma$ [11]. The analysis of the observed X-ray spectrum from the large Magellanic clouds and the Milky way in space experiments [12] can only put constraints on the mixing angle for different neutrino masses. However, even if the narrow X-ray line of approximately a few keV is observed in the further space experiments, the origin of it will be still questionable taking into account the conditions of the surrounding interstellar radiation. Moreover, galactic observations can predict only the mass value of DM particles, however not their type and properties. Meanwhile, there are other candidates for WDM (majorons, gravitons, neutralinos etc.). Therefore, a discovery of keV sterile neutrinos in terrestrial laboratory experiments would strengthen the assumption that keV sterile neutrinos could be the origin of WDM.

An overview of different approaches relevant to the search for keV sterile neutrinos is given in [13]. The authors of [13] emphasize the unique possibility of β-decay in terrestrial laboratory conditions for this search. They have proposed a sophisticated experiment with full kinematic measurements for the initial nuclide, recoil ion and electron with the hope to observe anomalous events caused by sterile neutrinos. This idea is still awaiting its implementation.

The authors of [9] have demonstrated that the KATRIN [14] and the MARE [15] experiments on the determination of the neutrino mass from the β-decay of $^3$H and $^{187}$Re, respectively, can also be employed for the search for keV sterile neutrinos. Meanwhile, the re-analyzed data of the Mainz [16] and the Troitsk [17] experiments with $^3$H have revealed no contribution of sterile neutrinos with mass values between 2 and 200 eV to the shape of the differential β-spectra. The lower limit of this contribution has been estimated as $U_{e4}^2 \approx 10^{-2}$ for $m_s \approx 100\,\text{eV}$ [17].

Atomic orbital electron capture (EC) provides an alternative approach for the search for keV sterile neutrinos. EC offers some advantages over β-decay such as a discrete spectrum of EC instead of a continuous one in the case of β-decay. Since EC is a two body process and the atomic electron binding energies are discrete and well known, the total energies of the released neutrinos should also be discrete and can be well measured too. Thus, the energy spectrum will consist of a series of peaks attributed to $Q_{\nu_i} = Q_{EC} - B_i$, where $Q_{\nu_i}$ is the neutrino total energy, $Q_{EC}$ is the atomic mass difference of the initial and the final atoms, and $B_i$ is the $i$-orbital electron binding energy ($i=K,L,M,...$orbits). The discrete energy spectrum can be measured by CM [18, 19], which embodies all atomic de-excitations (characteristic X-rays, Auger and Coster–Kronig electrons, etc.) following the electron vacancy occupation as well as the nuclear recoil energy.

This paper aims to demonstrate a new approach to search for keV sterile neutrinos in atomic electron capture by a nucleus, which can be implemented in terrestrial laboratories. The sensitivity of this method is estimated and the nuclides which can be used for these experiments are suggested. This method can be considered complementary to the exploration of the continuous differential β-spectrum [9].

## 2. Sterile neutrino contribution to the shape of an electron capture spectrum.

The probability to capture the $i$-orbital electron by a nucleus on the assumption that sterile neutrinos exist can be written as

$$\lambda_i = \frac{G^2}{4\pi^2} C_i \left\{ (1-U_{e4}^2)|q_i|_{act} Q_i + U_{e4}^2 |q_i|_{st} Q_i \right\}, \tag{2}$$

where $G$ is the weak coupling constant; $C_i = |M|^2 \cdot |\psi_i|^2$ is the form-factor for an allowed transition with $|M|^2$ being the squared nuclear matrix element; $|\psi_i|^2$ is the electron density at the origin for a point like nucleus or the average of the electron density over the nuclear volume for an extended nucleus. Next, $|q_i|_{act} = \sqrt{Q_i^2 - m_{act}^2}$ and $|q_i|_{st} = \sqrt{Q_i^2 - m_{st}^2}$ are the momentum of active and sterile neutrinos, respectively; $Q_i = Q_{EC} - B_i$ is the total active neutrino energy with the atomic mass difference $Q_{EC}$, denoted in the following as $Q$. Thereby with the assumption that $m_{st} \gg m_{act} \approx 0$ we derive to the following formula:

$$\lambda_i = \frac{G^2}{4\pi^2} C_i \left\{ (1 - U_{e4}^2)(Q - B_i)^2 + U_{e4}^2 (Q - B_i) \sqrt{(Q - B_i)^2 - m_s^2} \right\}. \qquad (3)$$

The energy values in eqs. (2) and (3) are in the units of $m_e c^2$. We do not consider here the higher order effects, which can be neglected in the keV scale (see for details [20, 21]).

The main idea of the proposed experiment is to undertake precision measurements of the mass differences in the electron capture process and independently to measure the atomic de-excitation spectrum, which is a discrete series of peaks around $B_i$ values. The acquired number of capture events in the peaks can be compared with the probability to capture the $i$-orbital electron by a nucleus on the assumption of the existence of only active neutrinos (by taking $U_{e4}^2 = 0$ and $m_s = 0$):

$$\lambda_i = \frac{G^2}{4\pi^2} C_i (Q - B_i)^2. \qquad (4)$$

Thus, the measure of the experiment is the ratio of the probabilities for the two different $i$- and $j$-captures ($i, j \equiv K, L$, etc.). It can be done by taking the ratio of numbers of counts $N_i/N_j$ under correspondent peaks in the spectrum measured by MC. Then, we have to compare this experimental value with the theoretical ratio:

$$\left( \frac{\lambda_i}{\lambda_j} \right)_{st} = \left( \frac{\lambda_i}{\lambda_j} \right)_{act} \cdot \frac{U_{e4}^2 \left( \mathrm{H}\left[ (Q - B_i) - m_s \right] \cdot \sqrt{1 - m_s^2/(Q - B_i)^2} - 1 \right) + 1}{U_{e4}^2 \left( \mathrm{H}\left[ (Q - B_j) - m_s \right] \cdot \sqrt{1 - m_s^2/(Q - B_j)^2} - 1 \right) + 1}, \qquad (5)$$

where 
$$\left( \frac{\lambda_i}{\lambda_j} \right)_{act} = \frac{|\psi_i|^2}{|\psi_j|^2} \cdot \frac{(Q - B_i)^2}{(Q - B_j)^2}. \qquad (6)$$

$\mathrm{H}\left[ (Q - B_i) - m_s \right]$ is the Heaviside step function, which is equal to 1 when $m_s \leq (Q - B_i)$ and 0 when $m_s > (Q - B_i)$.

In eqs. (5) and (6) $i$ and $j$ stand for different combinations of the allowed electron orbits $K$, $L$, $M$, etc. As can be seen from eq. (5), the sensitivity to $U_{e4}^2$ depends on the relation of the sterile neutrino mass $m_s$ to the difference between $Q$ and $B_i$.

The electron wave functions can be calculated only with a certain non-vanishing uncertainty. In order to exclude this origin of uncertainty the same ratios for the probabilities in different isotopes of the same chemical element must be compared. In this case the influence of the electron wave functions is cancelled to large extend (see section 3 and Table 2) and the sensitivity to $U_{e4}^2$ can be considerably increased. The capture probability ratios for the isotope numbers $1$ and $2$ can be written as

$$\zeta_{st} \equiv \frac{(\lambda_i/\lambda_j)_1}{(\lambda_i/\lambda_j)_2} = \zeta_{act} \frac{\left[ 1 - U_{e4}^2 (1 - \omega_{i1}) \right] \left[ 1 - U_{e4}^2 (1 - \omega_{j2}) \right]}{\left[ 1 - U_{e4}^2 (1 - \omega_{i2}) \right] \left[ 1 - U_{e4}^2 (1 - \omega_{j1}) \right]}, \qquad (7)$$

where 
$$\zeta_{act} = \left[ \frac{(Q_1 - B_i)(Q_2 - B_j)}{(Q_2 - B_i)(Q_1 - B_j)} \right]^2, \qquad (8)$$

and 
$$\omega_{lk} \equiv \mathrm{H}\left[ (Q_k - B_l) - m_s \right] \cdot \sqrt{1 - \left( \frac{m_s}{Q_k - B_l} \right)^2}.$$

Here $k = 1, 2$ is the number of the isotope and $l = i, j$ is the atomic electron orbit.

As can be seen from eqs. (7) and (8) the sensitivity to the sterile neutrino contribution with mass value $m_s$ depends on the accurate determination of atomic mass differences $Q_{1,2}$ for both isotopes. We assume that the electron binding energies $B_{i,j}$ are known with a precision better than $\approx 1$eV [23].

In order to be sure that the sterile neutrino contribution can be sufficiently large for both isotopes the Q-values $Q_1$ and $Q_2$ should be very similar. However, if one of them, e.g., $Q_2$, is much larger than the innermost orbit binding energy: $Q_2 \gg B_i$, then the contribution of sterile neutrinos to the shape of the atomic de-excitation spectrum (see section 3) of the second isotope can be neglected. This simplifies eq. (7) and gives:

$$U_{e4}^2 = \frac{\zeta_{\exp} - \zeta_{\mathrm{act}}}{\zeta_{\mathrm{act}}(1-\omega_{j1}) - \zeta_{\mathrm{act}}(1-\omega_{i1})}. \tag{9}$$

Here again the sensitivity to the sterile neutrino contribution in the atomic de-excitation spectrum depends on the accurate knowledge of the capture probability ratios $\zeta_{\exp}$ addressed by MC and precise values of $Q_{1,2}$, which can be measured by PT-MS (see section 5).

It is worthwhile to emphasize that eqs. (7) and (9) do not depend on the nuclear matrix elements and the atomic electron wave functions. This is a great advantage and an attractive issue for the sterile neutrino search in the electron capture sector. The estimations of the sterile neutrino contribution in electron capture with eqs. (5)–(8) are given below for several most promising EC transitions.

## 3. Nuclides relevant for the assessment of the keV sterile neutrino contribution in electron capture.

The physics parameters needed to determine the keV-sterile neutrino contribution $U_{e4}^2$ to the capture probability are the atomic mass difference $Q$ of the capture partners, the binding energies of the atomic electrons $B_i$, and the electron density at the origin $|\psi_i|^2$ (see eqs. (5)–(8)). If the information is deduced from the microcalorimetric spectrum of only a single transition (see eqs. (5) and (6)), then the electron wave functions $\psi_i$ should be calculated with the corresponding corrections [20]. The experimental sensitivity for the observation of sterile neutrinos depends on the precision of the mentioned parameters and on the statistics, i.e., the number of counts under the peaks in the microcalorimetric spectrum.

As can be seen from eq. (5) the highest sensitivity to the sterile neutrino mixing can be reached when the sterile neutrino mass value is close to the difference between $Q$ and $B_K$. At the same time, for heavy nuclides $m_s \approx Q - B_L$, $Q - B_M$, etc. must be fulfilled. This requires $Q < 100$ keV for the atomic mass difference. Only very few EC transitions fulfill these requirements relevant for the search for keV sterile neutrinos in the region 0.5 keV$< m_s <$100 keV; they are shown in Table 1. A similar list of nuclides was already considered in [8] in connection to the 17 keV-neutrinos.

Table 1. Nuclides with relevant energy balance for the search for the keV sterile neutrinos in the electron capture process (on the suitability of specific cases see section 4).

| Nuclide | $T_{1/2}$ | EC-transition | $Q$ (keV) [22] | $B_i$ (keV) [23] | $B_j$ (keV) [23] | $|\psi_i|^2/|\psi_j|^2$ | $Q-B_i$ (keV) |
|---|---|---|---|---|---|---|---|
| $^{123}$Te | $>2 \cdot 10^{15}$ y | ? | 52.7(16) | K: 30.4912(3) | $L_I$: 4.9392(3) | 7.833 | 22.2 |
| $^{157}$Tb | 71 y | $3/2^+ \to 3/2^-$ | 60.04(30) | K: 50.2391(5) | $L_I$: 8.3756(5) | 7.124 | 9.76 |
| $^{163}$Ho | 4570 y | $7/2^- \to 5/2^-$ | 2.555(16) | $M_I$: 2.0468(5) | $N_I$: 0.4163(5) | 4.151 | 0.51 |
| $^{179}$Ta | 1.82 y | $7/2^+ \to 9/2^+$ | 105.6(4) | K: 65.3508(6) | $L_I$: 11.2707(4) | 6.711 | 40.2 |
| $^{193}$Pt | 50 y | $1/2^- \to 3/2^+$ | 56.63(30) | $L_I$: 13.4185(3) | $M_I$: 3.1737(17) | 4.077 | 43.2 |
| $^{202}$Pb | 52 ky | $0^+ \to 2^-$ | 46(14) | $L_I$: 15.3467(4) | $M_I$: 3.7041(4) | 4.036 | 30.7 |
| $^{205}$Pb | 13 My | $5/2^- \to 1/2^+$ | 50.6(5) | $L_I$: 15.3467(4) | $M_I$: 3.7041(4) | 4.036 | 35.3 |
| $^{235}$Np | 396 d | $5/2^+ \to 7/2^-$ | 124.2(9) | K: 115.6061(16) | $L_I$: 21.7574(3) | 5.587 | 8.6 |

As can be seen from Table 1, the presented nuclides cover a rather wide region of sterile neutrino mass values. If the sterile neutrino mass obeys to the inequality $m_s \geq Q - B_i$, then it contributes to the capture only from the higher orbits $j$ ( $j > i$ ).

The signature for the existence of keV sterile neutrinos would be an at least 3σ deviation of the ratios (given in eq. (5)) from the expected ones on the assumption of no existence of keV sterile neutrinos (eq. (6)). On this assumption, the smallest value of the mixing matrix element $U_{e4}^2$ can be derived from eqs. (5) and (6) for a certain neutrino mass value. This analysis has been done for one capture pair of different chemical elements, as well as for two capture pairs of the same element, and the outcome of this analysis is presented below.

*One capture pair.*

Eqs. (5) and (6) applied to the nuclides from Table 1 yield that it is sufficient to acquire statistics with a decay rate of 10 atom/s for one year under the experimental conditions described in section 4, in order to reach a sensitivity of $10^{-5}$ to the sterile neutrino contribution. Figure 1 shows a behavior of the contribution curve in dependence on the sterile neutrino mass value.

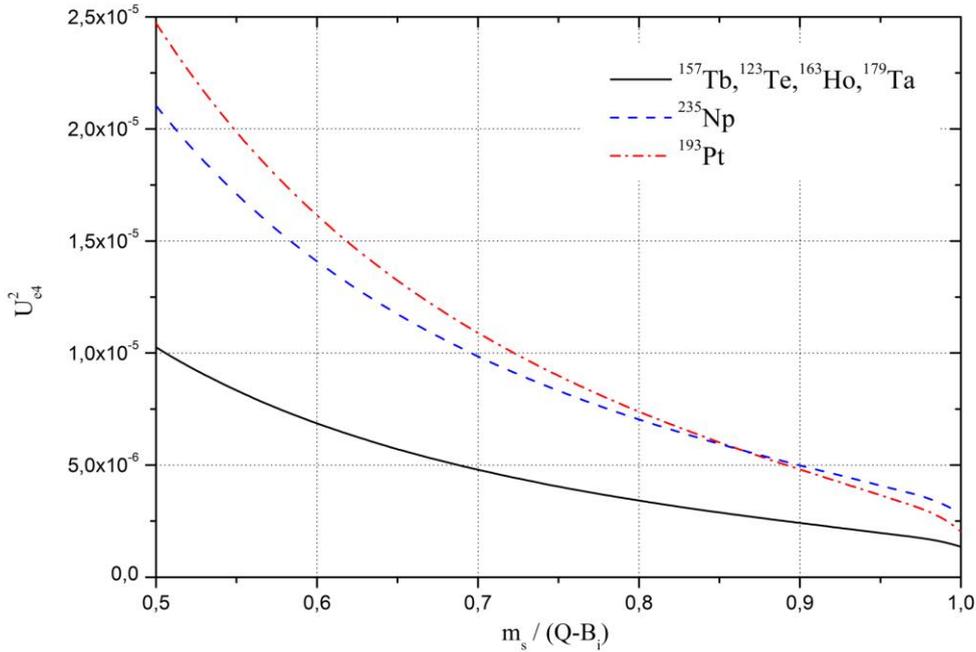

Fig. 1. Sensitivity to the mixing matrix element $U_{e4}^2$ which can be reached after one-year data acquisition with a decay rate of 10 atom/s as a function of the sterile neutrino mass for different nuclides listed in Table 1. The values of $U_{e4}^2$ for $^{123}$Te, $^{157}$Tb, $^{163}$Ho and $^{179}$Ta have such small differences that they cannot be resolved in the figure. The values $B_i$ correspond to the higher energy peak in the EC spectrum; $Q$–$B_i$ can be found in the eighth column of Table 1.

However, in these calculations we did not take into consideration the uncertainties for the quantities which enter in eqs. (5) and (6). In order to estimate the minimal sterile neutrino mixing matrix element $U_{e4}^2$ for a certain $m_s$, which can be deduced within the 3σ confidence level from the experimental data, the solution for the next equation should be found:

$$\left| (\lambda_i / \lambda_j)_{act} - (\lambda_i / \lambda_j)_{st} \right| = 3 \cdot \sqrt{\sum_k \left( \frac{\partial (\lambda_i / \lambda_j)_{act}}{\partial x_k} \delta x_k \right)^2} + 3 \cdot \sqrt{\frac{N_i^2}{N_j^3} + \frac{N_i}{N_j^2}} \ . \qquad (10)$$

Here $x_k = Q, B_i, B_j$; $N_i$ and $N_j$ – are the number of counts under the correspondent peaks in the experimental spectrum; expressions for $(\lambda_i / \lambda_j)_{st}, (\lambda_i / \lambda_j)_{act}$ have to be taken from eqs. (5) and (6). The so-lution of equation (10) for a certain mass will be indicated as $U^2_{e4\,min}$. Thereby, solving this equation for each mass of sterile neutrinos $m_s$ the dependence of $U^2_{e4\,min}$ on $m_s$ can be plotted. As can be seen from eq. (10) the smaller $\delta x_k$ and the larger $N_{i,j}$, the higher the sensitivity for $U^2_{e4}$. The binding energies $B_i$, $B_j$ are known with uncertainties of typically less than 1 eV [23].

Figures 2–6 show the values $U^2_{e4\,min}$ calculated with eq. (10) as a function of the sterile neutrino mass $m_s$ for several uncertainties of the $Q$-value of the EC transitions $^{123}$Te→$^{123}$Sb, $^{163}$Ho→$^{163}$Dy, $^{179}$Ta→$^{179}$Hf, $^{193}$Pt→$^{193}$Ir and $^{235}$Np→$^{235}$U respectively.

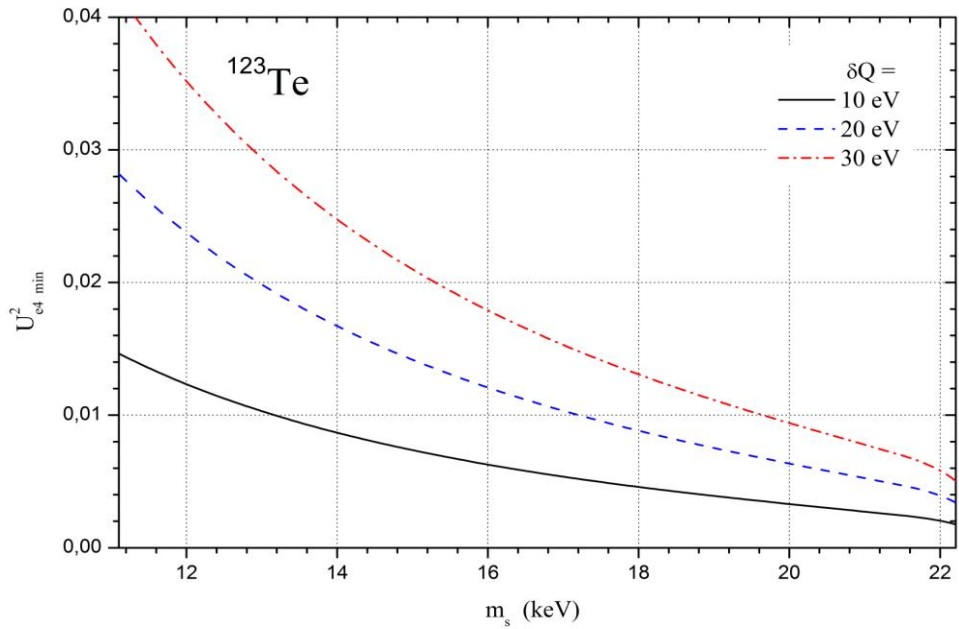

Fig. 2. Behavior of the minimal sterile neutrino mixing matrix element $U^2_{e4\,min}$ deduced within the 3σ confidence level from the experimental data of the capture-ratio probability value for $^{123}$Te in dependence on sterile neutrino mass values. The curves are estimated under the assumption that the uncertainties of the atomic wave functions are negligible, i.e. $\delta|\psi_i|^2 = 0$, and the uncertainties of the absolute atomic mass difference $\delta Q$ are 10, 20, and 30 eV, respectively.

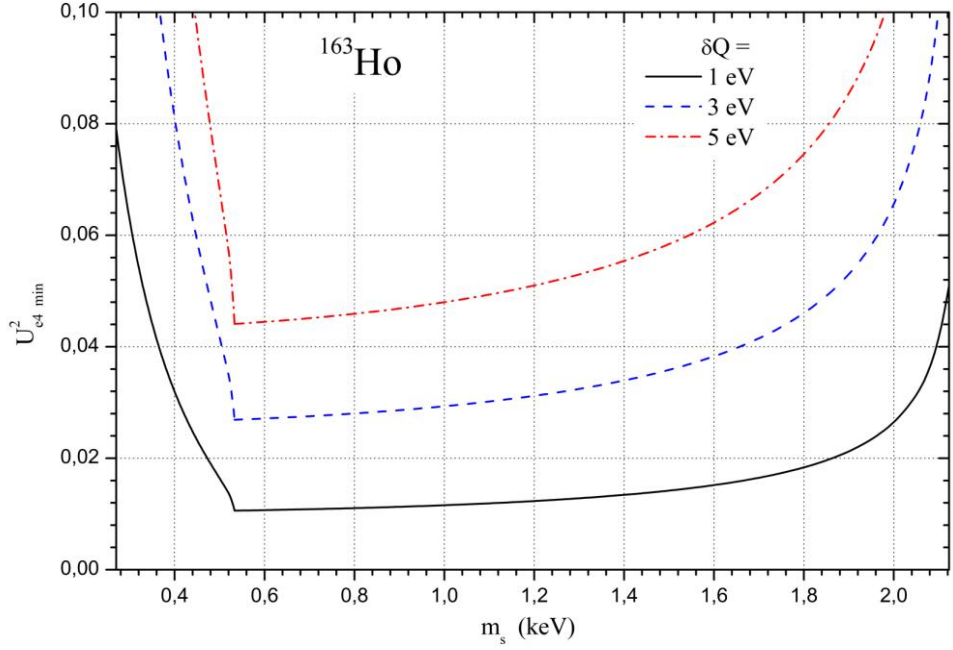

Fig. 3. The same as in Fig. 2 for the case of $^{163}$Ho. Note that the uncertainties of $\delta Q$ are now 1, 3 and 5 eV, respectively, and the range of $m_s$ is different.

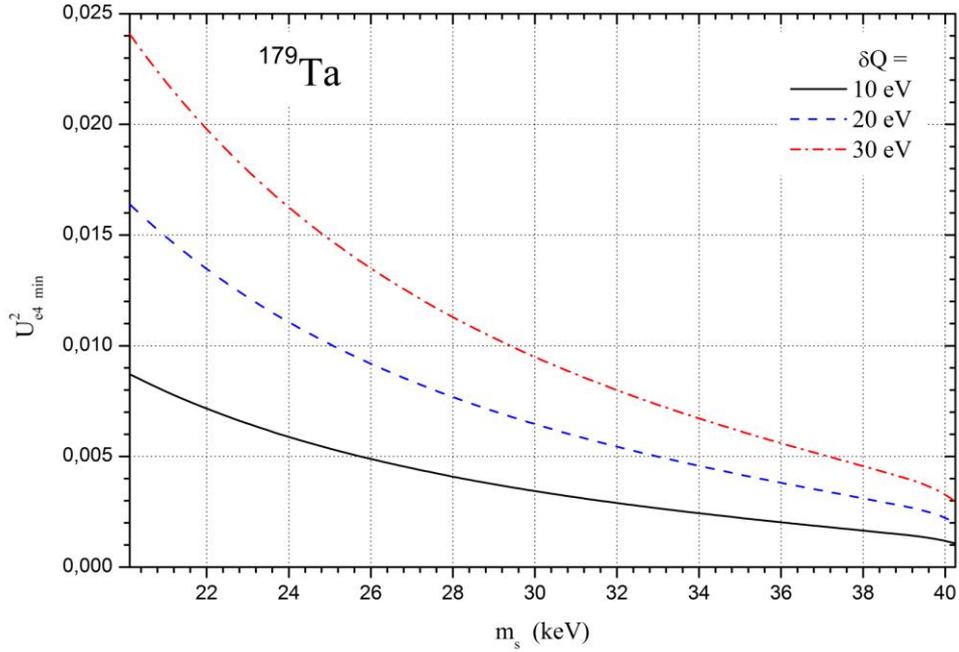

Fig. 4. The same as in Fig. 2 for the case of $^{179}$Ta. Note, however, that the range of sterile neutrino masses is different.

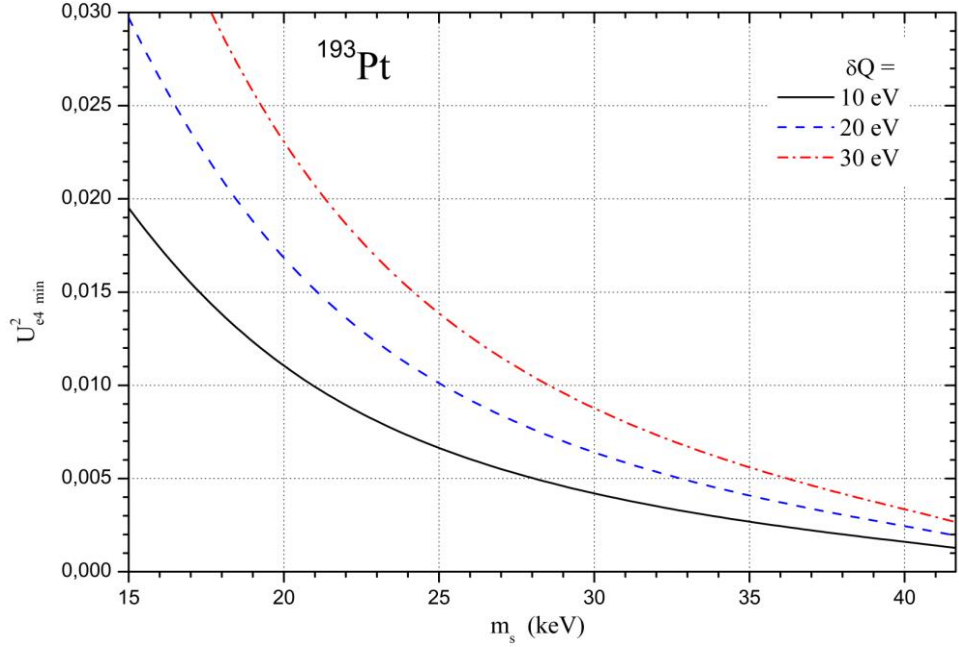

Fig. 5. The same as Fig. 2 for the case of $^{193}$Pt, but with a different scale for the sterile neutrino masses.

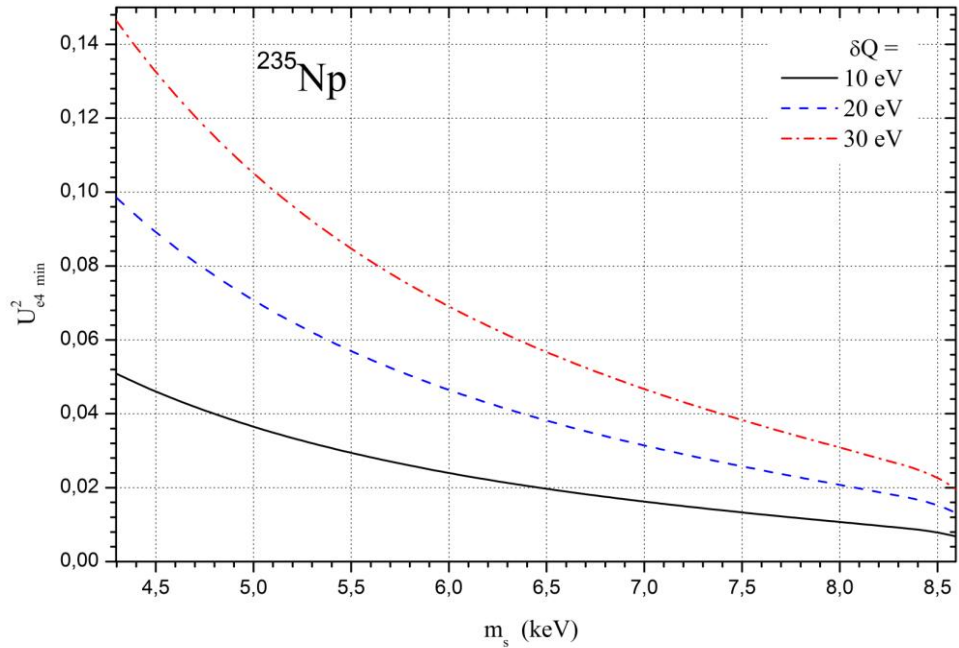

Fig. 6. The same as Fig. 2 for the case of $^{235}$Np, but with a different scale for sterile neutrino masses.

The lowest contribution of $U_{e4}^2$ on the level of only a few percent can be observed in the EC of the nuclides, if the neutrino mass is close to the corresponding $(Q–B_i)$-value. $B_i$ corresponds to the higher energy peak in the EC spectrum (usually the K-line, but for $^{163}$Ho is the $M_I$ line)

The $^{163}$Ho–$^{163}$Dy transition has the advantage that, in addition to the sensitivity to the small sterile neutrino mass, it is an allowed weak transition ($7/2^-\to 5/2^-$) and is currently under careful investigation by means of CM [24]. Thus, some information on the mixing of the sub-keV and keV sterile neutrinos can be deduced in the first place (see section 4). Since the $Q$-value of the EC in $^{163}$Ho is very small (see Table 1), the K- and L-capture are forbidden. In the calorimetric spectrum only $M_I$, $M_{II}$, $N_I$, $N_{II}$ and $O_I$ peaks have already been seen.

The nuclide $^{163}$Ho can be accumulated at the ISOLDE-facility at CERN [25] or by irradiating $^{162}$Er in nuclear reactors in a reasonable quantity in order to reach the sensitivity shown in Fig. 3.

Actually the ratios of the wave functions squared $\Psi \equiv |\psi_i|^2 / |\psi_j|^2$ can be presently estimated with a relative uncertainty $\delta\Psi/\Psi$ of about 0.01 [20]. This value puts substantial constraints on the precision for $(\lambda_i/\lambda_j)_{act}$. For this reason, all values $U^2_{e4\,min}$ in figures 2 – 6 become approximately five times bigger. In order to exclude the origin of this uncertainty the same ratios for probabilities in different isotopes of the same chemical element must be compared.

*Two capture pairs in different isotopes of the same chemical element.*

If the capture ratios are measured for different isotopes of the same chemical element, the atomic parts of the probability and the nuclear matrix elements are the same and are hence cancelled. The electron capture probability from the atomic orbit *nlj* is proportional to the one-electron density at the nucleus, which in the case of the point-charge nucleus is $\rho_{nlj}(0) = |\Psi^2_{nlj}(0)|$. In the case of an extended nucleus, this density $\rho_{nlj}(r) = |\Psi^2_{nlj}(r)|$, being dependent on *r*, considerably varies within the nucleus. However, the change of the ratio of two different orbit densities with the same angular symmetry, $C(r) = \rho_{n_1 lj}(r)/\rho_{n_2 lj}(r)$, generally does not exceed 0.1% for the variation of *r* within the extended nucleus. Another uncertainty of the atomic electron density factor comes from the electron-electron correlation effect. It turns out, however, that this uncertainty is strongly reduced, if the ratio of the values of $C(r)$ for two different isotopes is considered. These statements are confirmed by the direct calculations of the core electron densities, as an example, for two isotopes of Pb, presented in Table 2. It can be seen from the table that, whereas the ratios $C(r)$ calculated at the zero point ($r = 0$) and at the root mean-square nuclear radius ($r = R_{rms}$) differ by slightly less than 0.1%, the ratio of the values of $C(r)$ calculated for two different isotopes of Pb deviates from each other by less than $10^{-5}$ in both simple hydrogen-like model and more elaborated Dirac-Fock method [26]. It means that the electron-electron interaction contribution to this ratio is almost completely cancelled.

Table 2. Density ratios $C(r)$ for two isotopes $^{205}$Pb and $^{202}$Pb at $r = 0$ and at $r = R_{rms}$ ($R^{205}_{rms} = 5.4828(15)$ fm, $R^{202}_{rms} = 5.4705(17)$ fm [27]). The Fermi model for the nuclear charge distribution is used.

| $C(r)$ | H-like wave functions | | | Dirac-Fock wave functions | | |
|---|---|---|---|---|---|---|
| | $^{205}$Pb | $^{202}$Pb | $\delta^*$ | $^{205}$Pb | $^{202}$Pb | $\delta^*$ |
| $\rho_{1s}(0)/\rho_{2s}(0)$ | 5.7639079 | 5.7638700 | $0.66\cdot10^{-5}$ | 6.4911984 | 6.4911569 | $0.64\cdot10^{-5}$ |
| $\rho_{1s}(R_{rms})/\rho_{2s}(R_{rms})$ | 5.7676369 | 5.7675907 | $0.80\cdot10^{-5}$ | 6.4952026 | 6.4951522 | $0.78\cdot10^{-5}$ |
| $\rho_{1s}(0)/\rho_{3s}(0)$ | 19.200512 | 19.200372 | $0.73\cdot10^{-5}$ | 28.058279 | 28.058083 | $0.70\cdot10^{-5}$ |
| $\rho_{1s}(R_{rms})/\rho_{3s}(R_{rms})$ | 19.215372 | 19.215199 | $0.90\cdot10^{-5}$ | 28.078454 | 28.078213 | $0.86\cdot10^{-5}$ |
| $\rho_{2p_{1/2}}(0)/\rho_{3p_{1/2}}(0)$ | 2.8323155 | 2.8323117 | $0.13\cdot10^{-5}$ | 3.8474125 | 3.8474085 | $0.10\cdot10^{-5}$ |
| $\rho_{2p_{1/2}}(R_{rms})/\rho_{3p_{1/2}}(R_{rms})$ | 2.8326675 | 2.8326630 | $0.16\cdot10^{-5}$ | 3.8477859 | 3.8477810 | $0.13\cdot10^{-5}$ |

*) $\delta = \left[ C^{205}(r)/C^{202}(r) \right] - 1$

One can distinguish between two possibilities: either sterile neutrinos contribute in both isotopes (see eq. (7)), or only in one isotope (see eq. (9)) and for another isotope this contribution can be neglected because of the fulfillment of the condition $m_s \gg Q - B_i$ for the innermost electron orbit in the atom.

Fortunately, both of these cases can be found in practice by choosing certain nuclides which can predominantly capture electrons. Examples are:
   a) $^{205}$Pb → $^{205}$Tl and $^{202}$Pb → $^{202}$Tl,
   b) $^{157}$Tb → $^{157}$Gd and $^{158}$Tb → $^{158}$Gd.
We will discuss them individually in the following.

### $^{205}$Pb→$^{205}$Tl and $^{202}$Pb→$^{202}$Tl

Both pairs of nuclides have nearly equal atomic mass differences: $^{205}$Pb→$^{205}$Tl with $Q = 50.6 \pm 0.5$ keV and $^{202}$Pb→$^{202}$Tl with $Q = 46 \pm 14$ keV [22]. In order to perform reliable simulations for the sterile neutrino contribution, both $Q$-values must be measured. Especially it concerns the $Q$-value of $^{202}$Pb→$^{202}$Tl whose uncertainty is by far too large. Both nuclides of this transition can be produced, e.g., at the ISOLDE-facility at CERN [25] and measured with PENTATRAP [28, 29]. The nuclide $^{205}$Pb can be produced in a sufficient quantity by neutron irradiation of stable $^{204}$Pb.

Fig. 7 shows the minimal value of the sterile neutrino contribution vs. the sterile neutrino mass value $m_s$, which can be deduced from the $L/M$ ratio by solving the following equation for $m_s$ for the electron capture in $^{202}$Pb and $^{205}$Pb:

$$\left|\zeta_{act} - \zeta_{st}\right| = 3 \cdot \sqrt{\sum_k \left(\frac{\partial \zeta_{act}}{\partial x_k} \cdot \delta x_k\right)^2} + 3 \cdot \sqrt{\sum_{l,k}\left(\frac{\partial N_{exp}}{\partial N_{lk}} \cdot \delta N_{lk}\right)^2}, \qquad (11)$$

where $x_k = Q_1, Q_2, B_i, B_j$; $N_{exp} \equiv (N_{i1}/N_{j1})/(N_{i2}/N_{j2})$, index $k = 1, 2$ means different nuclides; index $l = i, j$ indicates different atomic electron orbits.

The atomic mass difference accuracy is assumed to be $\delta Q_{1,2} = 10, 20$, and $30$ eV, while $\delta Q$ is still a subject to careful measurements by PT-MS. As can be seen from Fig. 7, the sensitivity to a possible contribution of sterile neutrinos in the capture process reaches a few percent for the sterile neutrino mass of 10 keV — the most interesting value for WDM. For bigger values of $m_s$ it can be even below 1%.

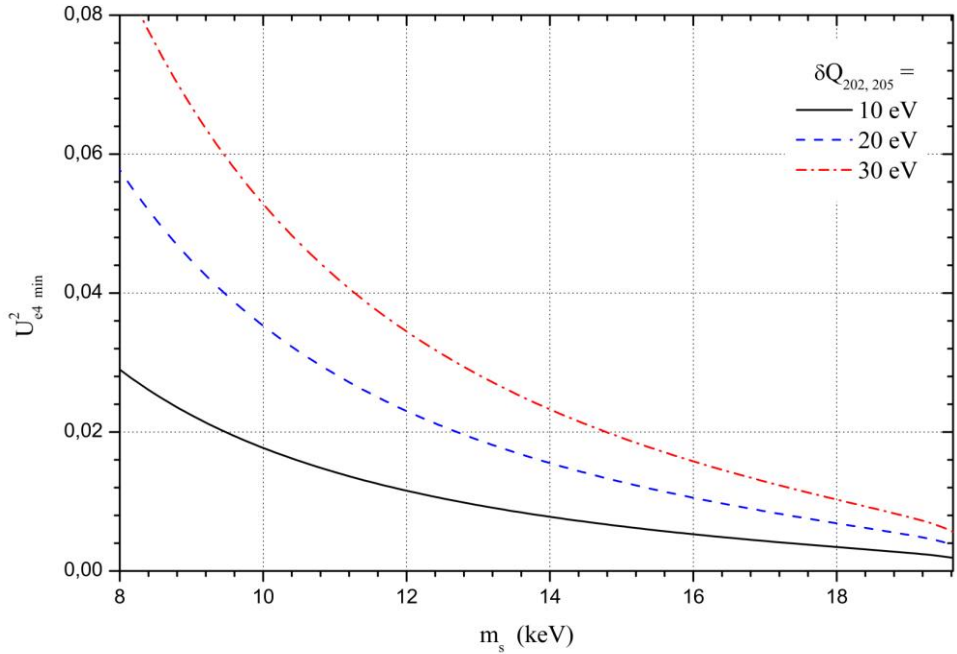

Fig. 7. The behavior of the minimal sterile neutrino mixing matrix element $U^2_{e4\ min}$ which can be deduced within the 3σ confidence level for $^{202}$Pb/$^{205}$Pb in dependence on sterile neutrino mass values. Note that neither the nuclear matrix element nor the atomic wave functions affect $U^2_{e4}$.

## $^{157}$Tb→$^{157}$Gd and $^{158}$Tb→$^{158}$Gd

The $Q$-value of the EC in $^{157}$Tb and the difference $Q - B_K$ (see Table 1) suit for the search for keV sterile neutrinos. Meanwhile, the strongest $K$-capture line in the $^{158}$Tb→$^{158}$Gd transition corresponding to $B_K$ = 50.239 keV is energetically far from the value of $Q$ = 197 keV for the transition to the ground state, or $Q$ = 178 keV for the other strong transition to the excited state of the stable daughter $^{158}$Gd [30]. Therefore, the contribution of keV sterile neutrinos in the calorimetric spectrum $^{158}$Tb→$^{158}$Gd should be very small because of the high energy differences $Q - B_K$ =147 and 128 keV for the excited state, respectively. However, see section 4 for details.

The simulated curves for the sterile neutrino contribution are given in Fig. 9. They have been estimated taking into account the ratios for K/L-capture probabilities in $^{157}$Tb and $^{158}$Tb (eq. (11)) and are shown as a function of the sterile neutrino mass value. Since the ratio of the K/L-capture probabilities for these two isotopes is no longer dependent on the uncertainty of the atomic part determination in the capture probabilities the expected precision of $U_{e4}^2$ can be much higher. It should depend mainly on the precision of the atomic mass difference measurement by means of PT-MS. The statistical uncertainty of the calorimetrically measured peaks contribute to the value $U_{e4\,\text{min}}^2$ in the figures 2–8 on the level of $10^{-4}$–$10^{-5}$. One day accumulation of both terbium isotopes with a production rate of $10^7$ atoms per second at the ISOLDE-facility [25] gives about $10^9$–$10^{10}$ decays per one year exposition time that will not produce considerable pile-up in the magnetic microcalorimeter.

Fig. 8 shows that a sensitivity of $10^{-2}$ for the sterile neutrino contribution in the experiments with $^{157}$Tb and $^{158}$Tb is achievable for a sterile neutrino mass of about 10 keV, and even less than 1% for $\delta Q$ = 1 eV.

Besides the physics parameters and production conditions, the possibility to accommodate the terbium isotopes in the absorber of micro-calorimeter also should be taken into account. As it is mentioned below in section 4, this problem was solved for Ho within the ECHo-project [24], and the possible "encapsulation" of Tb is under progress by this group.

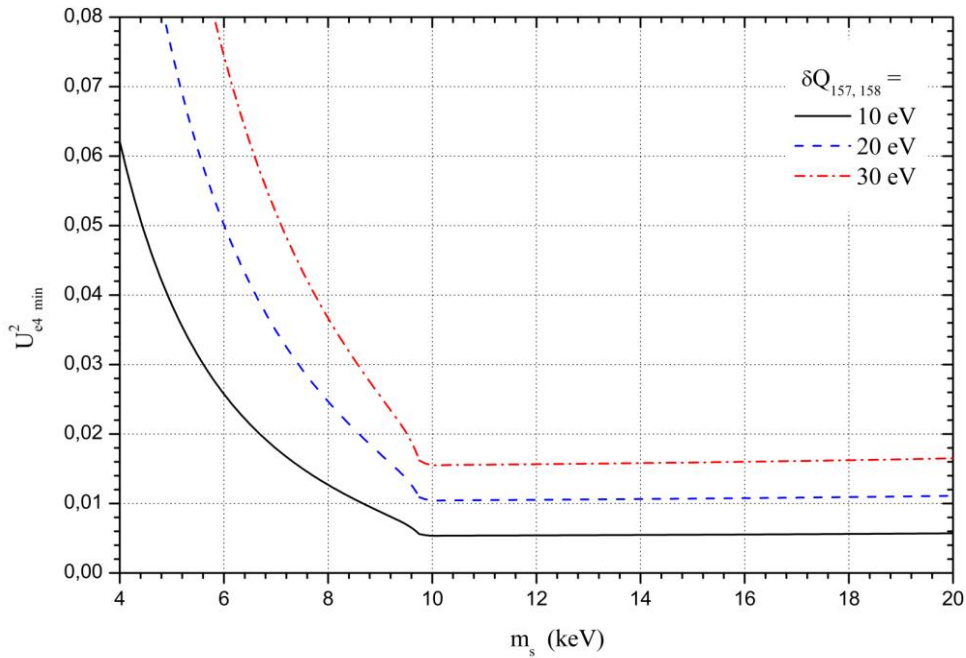

Fig. 8. Similar caption as for Fig. 7, but only for $^{157}$Tb/$^{158}$Tb.

As can be seen from Fig. 7 and Fig. 8, the experiments with two pairs of capturing nuclides of the same chemical element provide eventually a much higher sensitivity to a possible contribution of sterile neutrinos in the capture process than the investigation of only one separate pair. The latter has substantial constraints because of the 1%-uncertainty of the atomic wave function calculation. In the two-pair case, the sensitivity is not influenced by the nuclear and the atomic parts of the probability, and thus can reach for the above illustrated cases values of less than 1% for the sterile neutrino masses of up to 20 keV. Very precise mass measurements, however, have to be performed to reach this sensitivity.

It is worthwhile to note at the end of this section that the value $U_{e4}^2 \approx 10^{-2}$ was also claimed for the explanation of the neutrino anomaly in different neutrino experiments [2]. However, this anomaly has been explained by the existence of sterile neutrinos in the eV-region.

### 4. Atomic de-excitation spectra by cryogenic microcalorimetry.

The possibility to reach high sensitivity in the detection of the sterile neutrino contribution in EC spectra is strongly connected to the possibility to measure all the energy emitted in the decays minus the energy taken away by the neutrinos. Low temperature microcalorimeters are nowadays used to perform calorimetric measurements with very high quantum efficiency and very high energy resolution [31, 32, 33]. These detectors are able to measure not only the atomic de-excitation energy of the daughter atom, but also the nuclear recoil as well as possible de-excitation energy due to transitions from excited states of the chemical surrounding to their ground states. A drawback of the calorimetric method is that all the events occurring in the detector are measured. This means that the measured spectrum will have a source of intrinsic background that is the unresolved pile-up spectrum. In first approximation the fraction of unresolved pile-up events is given by the activity times the signal rise-time. For a given signal rise-time the unresolved pile-up fraction can be reduced by reducing the activity of the detector at the cost of requiring either a longer measuring time or more detectors. A discussion how to optimize the activity per pixel in order to make the effect of unresolved pile-up negligible is not given here, but for all the cases discussed it will be possible to reach this goal.

In this work we have considered low temperature metallic magnetic calorimeters (MMCs) [34]. They are energy dispersive low temperature detectors operated at temperatures below 100 mK. They make use of the calorimetric principle where the absorption of energy produces an increase of the detector temperature proportional to the deposited energy $\Delta E$ and to the inverse of the detector heat capacity $C_{\text{tot}}$. A paramagnetic temperature sensor, which resides in a small magnetic field, is tightly connected to the particle absorber and weakly connected to a thermal bath. The change of temperature leads to a change of magnetization of the sensor which can be read-out as a change of flux by a low-noise high-bandwidth dc-SQUID. The sensor material, presently used for MMCs, is a dilute alloy of erbium in gold, Au:Er. The concentration of erbium ions in the sensor can be chosen to optimize the detector performance and usually varies between 200 ppm and 800 ppm. MMCs can be optimized to be used in different energy ranges. In [35] the optimization of detectors for x-rays below 20 keV and for x-rays below 200 keV has been discussed. In the first case an energy resolution of $\Delta E_{\text{FWHM}} = 2$ eV at 6 keV and a pulse rise time $\tau_r = 90$ ns were measured while in the second case an energy resolution of $\Delta E_{\text{FWHM}} = 60$ eV at 60 keV and a rise time of time $\tau_r \approx 200$ ns. The energy resolution achieved by MMC detectors is among the best achieved worldwide while the rise-time is shorter compared to similar other detectors. This indicates that calorimetric measurements of EC spectra performed with MMCs will be less affected by the unresolved pile-up background.

To perform a calorimetric measurement of an EC spectrum, the source has to be:
• part of the sensitive volume of the detector to prevent the partial loss of energy;

• homogeneously distributed in the detector so that the response is position independent;

• completely contained in the detector to ensure a quantum efficiency for the emitted particles of 100%.

A reliable method to embed sources in fully micro-fabricated MMC detector is the ion-implantation. This method has already been employed in preparing detectors for the calorimetric measurement of $^{163}$Ho. The ion implantation process was performed at ISOLDE-CERN [36]. A detailed description of these detectors is given in [37]. In order to reach a quantum efficiency very close to 100% the $^{163}$Ho source was implanted onto a first half of the energy absorber, consisting of a $(190 \times 190 \times 5)$ μm$^3$ of Au over a reduced area of $160 \times 160$ μm$^2$. After the implantation process a second gold layer having the same dimension with that of the first one was deposited on top of the first one. In this way less than $10^{-7}$% of the emitted radiation might not be stopped in the detector. An energy reso-lution of $\Delta E_{\text{FWHM}} = 7.6$ eV at 6 keV was measured [38]. The results achieved by these detectors showed first of all that the ion implantation process can be used to embed the source in the detectors and second that the calorimetric measurement of the $^{163}$Ho EC spectrum can be performed with very high quantum efficiency. With MMC detector having ion-implanted $^{163}$Ho it will be possible to perform the experiment to investigate the effects of the sterile neutrino in the spectrum. This investigation will be one of the goal of the ECHo experiment [38] where a large number of MMCs detectors, of the order of $10^5$, using the microwave multiplexing scheme, will be measured in order to reach a statistics of $10^{14}$–$10^{16}$ events in the full energy spectrum.

MMC detectors will also be able to measure with high energy resolution, of about $\Delta E_{\text{FWHM}} = 60$ eV and high quantum efficiency the EC spectrum of $^{193}$Pt and $^{179}$Ta, for the nuclides to be employed in the "one capture pair" approach. Both $^{193}$Pt and $^{179}$Ta decay into stable nuclides, $^{193}$Ir and $^{179}$Hf, respectively. The detectors that can be used in these experiments will be similar to the ones discussed in [35] for an x-ray energy below 200 keV. The energy absorber will have an almost cubic form and the dimensions will be adjusted to reach a quantum efficiency close to 100% for the particle emitted in the decay. As in the case of $^{163}$Ho, the source will be ion-implanted in a reduced area in the center of the first part of the absorber. Both source can be produced and implanted at ISOLDE. In order to reach the relatively large statistics so that the statistical error will be negligible with respect to the systematic error the same technology developed for the ECHo experiment can be used.

The case of $^{235}$Np will be more difficult to study since it is part of a long decay chain. Since the MMC detectors will measure all the emitted energy in the absorber, a large background due to the decay of the other nuclides present in the chain is expected. This fact will reduce the sensitivity to identify the effect of the sterile neutrino.

The case of $^{123}$Te is also difficult to study by means of low temperature microcalorimeter due to the fact that the half-life is expected to be longer than $10^{15}$ years. In this case a relatively large mass of $^{123}$Te should be embedded in the detectors. Presently investigation on the EC in $^{123}$Te is performed within the CUORE experiment where low temperature macro-calorimeters are used [39, 40]. This experiment might be able to detect for the first time the EC in $^{123}$Te.

The discussion concerning the possibility of performing experiments to investigate two pairs of different isotopes of the same chemical element is more complicated. In the case of the pair $^{157}$Tb→$^{157}$Gd and $^{158}$Tb→$^{158}$Gd the calorimentric measurement of the $^{157}$Tb spectrum is possible and a preliminary experiment was performed at the Kirchhoff Institute for Physics. In this experiment the $^{157}$Tb was implanted at ISOLDE in gold absorbers of a MMC detector. A preliminary measurement showed that the calorimetric spectrum of $^{157}$Tb can be measured with good energy resolution and high quantum efficiency. The measurement of the energy spectrum of $^{158}$Tb is more complicated. There are two main reasons. First of all, $^{158}$Tb decays to $^{158}$Gd via EC with 83.4% and decays to $^{158}$Dy via beta decay with 16.6% probability. The second decay branch will produce a background to the EC spectrum. The second reason is the high end-point energy of the EC process, $Q = 1220$ keV. In this case a full calorimetric measurement will not be possible, but the detector can be designed to have a quantum efficiency close to 100% for energies slightly above the K-line value. In order to reach the expected sensitivity to the sterile neutrino a precise estimation of the background due to the decay $^{158}$Tb→$^{158}$Dy should be done as well as the study of possible systematic errors due to the not completely calorimetric measurement.

In the case of the pair $^{202}$Pb→$^{202}$Tl and $^{205}$Pb→$^{205}$Tl the calorimentric measurement of the $^{205}$Pb EC spectrum using MMC detectors is possible and a high quantum efficiency and a high energy resolution can be achieved as described for $^{193}$Pt and $^{179}$Ta. Slightly more complicated is the scenario for the calorimetric measurement of the $^{202}$Pb EC spectrum. The calorimetric measurement of this spectrum can be performed with detectors similar to the ones that will be developed for the study of the decay of $^{157}$Tb, but this measurement will be affected by an intrinsic background. A relatively negligible fraction of this background is due to the partial decay of $^{202}$Pb to the stable $^{198}$Hg through α-emission. The large fraction of the background is due to the decay of the $^{202}$Tl to the stable $^{202}$Hg through EC. In order to reach the expected sensitivity on the sterile neutrino, a precise estimation of the intrinsic background has to be performed.

## 5. High-precision $Q_{EC}$ measurements by Penning-trap mass spectrometry.

As mentioned in section 2, the sensitivity for the determination of the sterile neutrino contribution to the shape of the EC atomic de-excitation spectrum considerably depends on the uncertainty of the mass difference $Q$ of the capture partners. For the EC transitions considered in the following, uncertainties of about 10 eV and below are required. At present, only high-precision Penning-trap mass spectrometry (PT-MS) is capable of providing such uncertainties for a broad variety of nuclides.

A determination of the $Q$-value of a certain transition, i.e., the mass difference $Q/c^2 = M_i - M_f$ of the initial and final states of the transition, respectively, is performed in PT-MS by measuring the cyclotron frequencies of the initial and final ionic states of the transition in a strong static homogeneous magnetic field $B$ [41]. The cyclotron frequency of a charged particle with the mass $M_{ion}$ and charge state $q$ is given by $\nu_c = \frac{1}{2\pi} \cdot \frac{q}{M_{ion}} \cdot B$. In order to push the uncertainty of the $Q$-value determination down to an eV level, the ion must be confined to a well-localized volume within the homogeneous magnetic field for at least some seconds. This is achieved by a superposition of a static three-dimensional quadrupole electric field on the magnetic field such that an electrostatic potential well along the magnetic field lines is created. The presence of the electrostatic quadrupole field modifies the pure ion's cyclotron motion to three independent trap motions: modified cyclotron, magnetron and axial motions with the trap eigenfrequencies $\nu_+$, $\nu_-$ and $\nu_z$, respectively. The invariance theorem gives a simple relation between the pure cyclotron and the trap frequencies [42]: $\nu_c^2 = \nu_+^2 + \nu_-^2 + \nu_z^2$. Thus, the determination of the ion's cyclotron frequency is performed via a measurement of the ion's trap eigenfrequencies.

The most accurate mass measurements with light stable atoms and molecules in closed systems has been reached with a relative mass uncertainty of $10^{-11}$ or even below with the UW-PTMS (Washington, USA) [43, 44] and with the MIT-TRAP (Massachusetts, USA) [45]. The MIT-TRAP moved about 10 years ago to FSU (Florida, USA). Nevertheless, until now there have been no mass measurements with a similar uncertainty on unstable medium-heavy and heavy nuclides, to which all capture partners considered in this work belong.

In order to extend such precise mass measurements into the region of unstable heavy nuclides, a novel cryogenic Penning-trap mass spectrometer PENTATRAP (Fig. 9) is currently being under construction at the Max-Planck Institute for Nuclear Physics [28, 29].

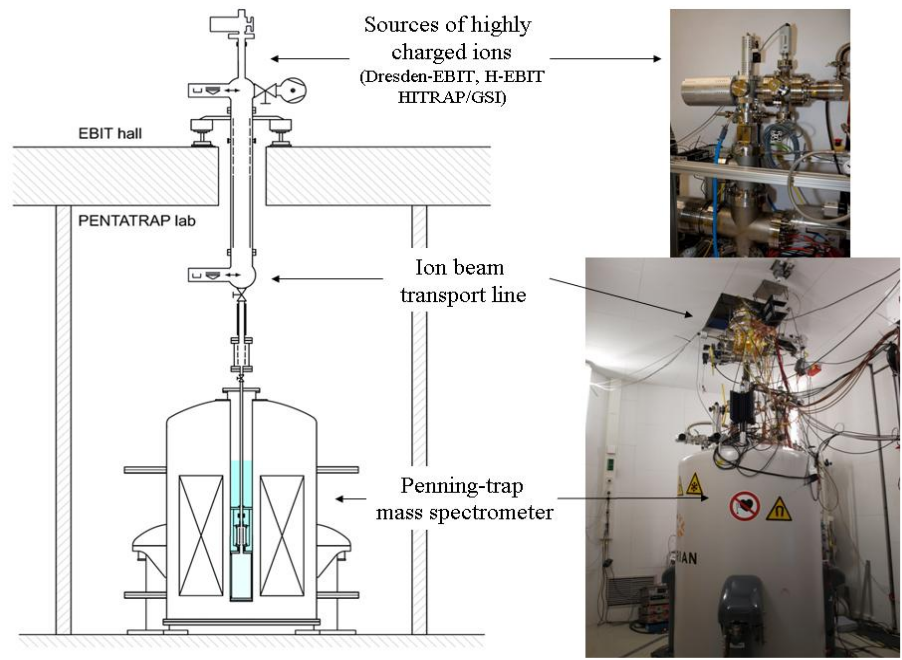

Fig. 9. Sketch and photo of the Penning-trap mass spectrometer PENTATRAP. For details see [28, 47]. In the upper right corner the EBIT is shown.

The uniqueness and complexity of this Penning-trap mass spectrometer is conditioned by the relative mass accuracy of $10^{-11}$. This can be achieved only by use of ions of interest in very high charge states, which will be produced with external electron-beam ion-trap sources, e.g., the commercial Dresden EBIT [48] and the H-EBIT [49]. Furthermore, the cyclotron frequencies of the initial and final states have to be measured simultaneously. For this, five cylindrical Penning traps will be employed and the novel cyclotron-frequency measurement technique described in [50] will be applied. Great care is taken to stabilize the magnetic field and the trap voltage. The effect of the environmental influence on the ion's motion in the traps is minimized by the stabilization of the temperature in the experimental room and by the screening of the magnet from stray electrical and magnetic fields. All these measures will allow a determination of the $Q$-values of the capture partners on the desirable level.

### 6. Conclusions.

The electron-capture sector can provide necessary conditions for the search for sterile neutrinos in the keV mass region. For this purpose two independent experiments should be undertaken: a measurement of the atomic mass difference ($Q$-values of electron capture) and a calorimetric measurement of the atomic de-excitation spectrum. Both domains have been dramatically developed over the last years: $Q$-values can already be measured with an uncertainty below 100 eV with conventional on-line Penning-Trap Mass Spectrometry of radioactive nuclides. The novel Penning-trap mass spectrometer PENTATRAP aims to push the achievable uncertainty down to 1 eV in the near future. Meanwhile, the considerable progress achieved in cryogenic magnetic microcalorimetry allows for measurements of different samples of nuclides with vanishingly small background. The existence of sterile neutrinos can be revealed by observing the shape distortion of the calorimetric spectrum, and, since the spectrum is discrete, can be derived from the ratios of the areas under the peaks corresponding to the different capture channels. The unknown nuclear matrix elements have to be cancelled in these ratios. If the ratios are measured for two different atomic pairs of the same chemical element, the atomic part of the transition matrix element should be also cancelled. In this case the experiment becomes more sensitive to the sterile neutrino contribution. Among the candidates for such measurements of the capture ratios the pairs $^{157}$Tb/$^{158}$Tb and $^{202}$Pb/$^{205}$Pb can be marked out. Calculations show that in dependence on the sterile neutrino mass values the contribution of sterile neutrinos in the neutrino mixing matrix from a few to less than one percent can be deduced from the feasible experimental data. This value is still much higher than expected for explanation of Warm Dark Matter, which, however, was simulated with model dependent cosmological assumptions.

### Acknowledgements

This work was supported by the BMBF (project 10/048), EMMI, Russian RFBR (grant No. 13-02-00630) and Russian Minobrnauki (project 2.2). P.F and Yu.N. would like to thank Max-Planck Institute for Nuclear Physics in Heidelberg for warm hospitality. We are indebted to E. Akhmedov for fruitful discussions.